\documentclass[aps,prb,letter,twocolumn,groupedaddress,floatfix]{revtex4}
\usepackage{graphicx}
\usepackage{epsfig}
\usepackage{float}
\usepackage{multirow}

\newcommand{\be}{\begin{equation}}
\newcommand{\ee}{\end{equation}}
\begin{document}
\title{Comparison of the Electronic Structures of Hydrated and Unhydrated Na$_x$CoO$_2$: The Effect of H$_2$O}
\author{M. D. Johannes, D. J. Singh}

\affiliation{Code 6391, Naval Research Laboratory, Washington, D.C. 20375}

\begin{abstract}
 	We report electronic structure calculations within density functional theory for the hydrated superconductor Na$_{1/3}$CoO$_2$1.33H$_2$O
and compare the results with the parent compound Na$_{0.3}$CoO$_2$.  We find that the intercalation of water into the parent compound has
little effect on the Fermi surface outside of the predictable effects expansion, in particular increased two-dimensionality. This implies
an intimate connection between the electronic properties of the hydrated and unhydrated phases. \end{abstract}

\maketitle

\section{Introduction} Two important discoveries during the past year, the likely unconventional superconductivity of Na$_x$CoO$_2$$\cdot
y$H$_2$O ($x\sim$1/3, $y\sim$4/3) and the unusual magnetotransport properties of Na$_x$CoO$_2$ (x$\sim$2/3), have focussed attention on
these materials and the connection between them.\cite{BGL03} In fact, layered cobalt oxide materials have been the subject of considerable
fundamental and practical interest in the last year for several reasons. Li$_x$CoO$_2$ is an important cathode material for lithium
batteries. In that context, the interplay between the transition metal-oxygen chemistry, the Co mixed valence and magnetism are important
ingredients in the performance of the material. \cite{ceder} Layered cobaltates, $A_x$CoO$_2$ also form for $A$=Na and K, but in more
limited concentration ranges. \cite{VMJ74} Single crystals of Na$_x$CoO$_2$, with nominal $x$=0.5 were investigated by Terasaki and
co-workers. \cite{ITYS99} Remarkably, they found that even though the material is a good metal, it also has a large thermopower of
approximately 100 $\mu$V/K at room temperature. This was the first time that an oxide showed promise of matching the thermoelectric
performance of conventional heavily doped semiconductors for thermoelectric power conversion. Interest in modifications of this material to
minimize its thermal conductivity for thermoelectric applications led to the discovery that similar anomolously high thermopowers were
present in other materials with hole doped CoO$_2$ layers, especially so-called misfit compounds in which the intercalating Na is replaced
by more stable rocksalt like oxide blocks. \cite{SHSL+01, YMMO+02, TYKU02, MHAM+03b} This demonstrates that the exceptional thermopower of
metallic Na$_x$CoO$_2$ is not essentially related to the details of the intercalating layer.

Theoretical studies, stimulated by these discoveries, emphasized both the band-like nature of the
material, \cite{DJS00,RAJS02b} consistent with its good metallic properties, the proximity to magnetism, and possible renormalizations
related to a magnetic quantum critical point, \cite{DJS03} the proximity to charge ordering (which is seen at some doping levels),
\cite{JKK-L03} and possible strongly correlated electron physics \cite{GB}. Intriguingly, both band structure and strong correlated models
({\it i.e.} the Heikes model) are able to explain the high thermopowers. \cite{WKSM01,WKSM03} Recently, it was demonstrated by
magnetotransport measurements that the thermopower at $x\sim 0.68$ is strongly reduced in magnetic field with a universal scaling law
\cite{YWNSR+03} showing that spin entropy underlies the high thermopower and thus again emphasizing the role of magnetic fluctuations in
the system as well as possible strong correlated electron physics. \cite{YWNSR+03,AMSH+03} Indeed, some of the misfit compounds are in fact
magnetic, with ferromagnetic ground states. \cite{ITTY+01} 

Takada and co-workers recently showed that Na$_x$CoO$_2$ can be readily hydrated
to form Na$_x$CoO$_2 \cdot y$H$_2$O. This material has the same CoO$_2$ layers, but with a considerably expanded $c$ axis, which
accomodates the intercalating water and Na. In this material, $x\sim 0.3$, is lower than the range readily formed in Na$_x$CoO$_2$.
Remarkably, Takada and co-workers found that Na$_x$CoO$_2 \cdot y$H$_2$O is a superconductor with $T_c \sim 5$K. \cite{KTHS+03,RES+03} The
nearness to magnetism and possible strong correlations immediately lead to suggestions of unconventional superconductivity in this
material, beginning with the discovery paper of Takada, as well as discussions of the role of water in producing the superconductivity.
Scenarios that have been advanced include no role at all, screening Na disorder, preventing competing charge ordered states, modifying the
doping level via unusual chemistry, enhanced two-dimensionality, and others. \cite{HSKT+03,GB03,BLJC+03,MLF+03,CH+03,MO+03,SPYL+03} 

Since superconductivity is fundamentally an instability of the metallic Fermi surface, a first step is to understand the relationship of the
electronic structure of Na$_x$CoO$_2 \cdot y$H$_2$O with its unhydrated parent Na$_x$CoO$_2$. Here we present density functional based
bandstructure calculations of the paramagnetic (non-spin polarized) compound using the linearized augmented planewave (LAPW) method
\cite{Wien2k,DJbook} for Na$_{1/3}$CoO$_2$$\cdot$4/3H$_2$O with both Na ions and water molecules explicitly included (no virtual crystal
approximation is made). Despite the probable presence of strong correlations in both hydrated and unhydrated compounds, we expect that, as is
the case with the high T$_C$ cuprates, an LDA bandstructure will give valuable and accurate information about the Fermi surface. ARPES
measurements on the parent compound of the superconducting cobaltate \cite{MZH+03,H-Y03} studied here already match well with previously
calculated (unhydrated) bandstructure results \cite{DJS00}.  We show that, from an electronic structure point of view, the hydrated and
unhydrated compounds are identical, aside from structural effects due to the expansion of the c-axis.

\vspace{0.1 in} 
\section{Hydrated Structure}

	There are a variety of ideas about the exact structure of the hydrated compound.  All reports \cite{JWL+03, JDJ+03, RJBCS+03,
RES+03,HSKT+03} indicate that it belongs to the hexagonal symmetry P6$_3$/mmc (\# 194), but refinements of the water molecule positions,
Na ion positions, and apical oxygen heights vary.  Lynn {\it et al} \cite{JWL+03} find that the Na ions are displaced compared to the
unhydrated parent compound and are surrounded by H$_2$O molecules with basically the same structure as D$_2$O ice.  However, other neutron
diffraction studies show that \cite{TE}, even below the freezing point of water, there may be no static position for the water molecule as
a whole, emphasizing disorder. 

\begin{center} \begin{figure}[h] \hspace{0.5 in} \includegraphics[width = 2.0 in]{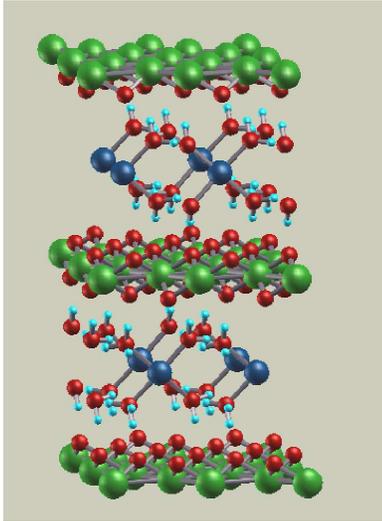} \caption{(Color online) The tripled, hydrated
structure corresponding to the superconductor.  The three planes of distorted octahedra are Co-O, the intercalated Na ions are 
drawn with bonds to the water molecules to emphasize the four-fold coordination. H ions point generally toward the Co-O planes.}
\end{figure}
\end{center}

The structure we used for our calculation was based on the neutron and x-ray diffraction results of Ref. \onlinecite{JDJ+03} shown in Fig. 1,
employing the observed symmetry sites, lattice constants, and composition.  Jorgensen {\it et al} obtained their model by systematically
removing ions/molecules from the partially occupied Na (6$h$) and H$_2$O (24$l$) sites according to coordination and bonding rules until the
observed proportions were obtained.  They held bond angles to 109$^{\circ}$ and O-H bond distances to 0.99 \AA {\i.e.} the water was assumed to
maintain its molecular characteristics.  To employ this structural configuration, we tripled the formula unit of the parent compound
Na$_{2x}$Co$_2$O$_4$ (already doubled to account for both Co-O planes), expanded the c-lattice to its reported value \cite{JDJ+03} of 37.1235
a.u. and added four H$_2$O molecules for each Na ion, resulting in a formula unit with integer values of all constituent atoms:  
Na$_2$Co$_6$O$_{12}$8H$_2$O.  This eliminated the need for the virtual crystal approximation while maintaining the proper proportion of each
ion, allowing us to take the possible effects of Na ordering into account, which we find {\it a posteriori} to be unimportant for the CoO$_2$
derived electronic structure, based on comparison with previous virtual crystal results.  Since the Na and H$_2$O sites are only partially
occupied in the P6$_3$/mmc symmetry, we required a different space group for computation.  Our structure has the considerably lowered P2$_1$/m
symmetry (\#11), but remains pseudohexagonal with lattice vectors of length $\sqrt{3}$a such that the planar area of the unit cell is tripled.  
In other words, we include the local structure and coordination, but not long range disorder in the Na H$_2$O layers, yielding a lower average
symmetry.  However, electronic structure around the Fermi level is hexagonal to a high precision, indicating that scattering from disorder in
the Na H$_2$O layers is weak.  This may be important for superconductivity considering the pair breaking effect of scattering in unconventional
superconductors.  Since no data is available for the exact orientation of the water molecules, they were oriented with H ions pointing away
from the Na ion and toward the Co-O plane, as suggested by various experimental observations \cite{JDJ+03,JWL+03,TE}. One hydrogen ion was
brought as near as possible to an oxygen in the Co-O plane to model possible O-H bonding.  The position of the other hydrogen was then
established by respecting the bond angles and lengths as described above.  We found that the electronic structure aspects that we discuss in
subsequent sections were insensitive to our particular choices for the water molecules. The apical O height in the presence of water was then
resolved to its optimal height of 1.86 a.u. above the Co plane.  The present local density approximation calculations were done using the LAPW
method as implemented in the WIEN code with well converged basis sets employing an Rkmax of 4.16, sphere radii of 1.86 (Co), 1.6 (O), 2.0 (Na),
and 1.0 and 0.88 for the O and H respectively of water.  The water molecules were treated using LAPW basis functions, whereas all other ions
were treated with an APW + LO basis set.  Additional local orbitals were added for Co and Na p-states and O s-states.

\section{Bandstructure and Fermi Surface}
	The most important bands in the conventional hexagonal Brillouin zone (BZ) of the unhydrated parent compound are four E$_g$' and
two A$_{1g}$ Co-derived bands near the Fermi energy. Our expanded hexagonal unit cell results in a BZ one third the volume of the original
and rotated by an angle of 30$^{\circ}$. The rotation and compression of the original BZ necessitates a double downfolding process as
illustrated in Fig. 2, and needs to be remembered when comparing our band structure and prior results. \cite{DJS00,JKK-L03}. 

\begin{figure} [h] \includegraphics[width = 1.2 in]{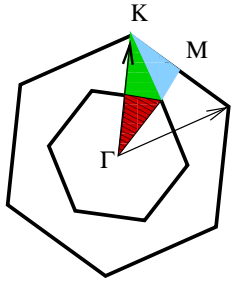} \caption{(Color online) The bands along the $\Gamma$-$M$ in the small zone are formed
by folding the upper shaded triangle down onto the adjacent one and then folding again into the irreducible Brillouin zone. The symmetry points
marked are those of the {\it larger} (unhydrated compound) zone, those of 
the smaller zone are easily identifiable by analogy} \end{figure}

To clarify the similarities and differences between the hydrated and unhydrated structures, we performed a second calculation in a similar
unit cell, neglecting the water molecules. We found that artificially expanding the c-axis with a vacuum produced unphysical
and highly dispersive bands.  The c-axis in our comparison calculation was fixed at its lower unhydrated value of 20.4280 a.u. for this
reason. 

Fig. 3 shows both bandstructures on the same energy scale, each centered around its respective Fermi energy.  The difference in c-axis
parameter is reflected in the $\Gamma$ - $A$ distance which is nearly twice as big in the parent compound.  An inspection of the bands crossing
and just above the Fermi energy reveals a somewhat greater splitting in the unhydrated compound than in the superconducting compound, but a
nearly identical overall band dispersion.  Bands containing water character are determined to be at least 0.2 Ryd below the Fermi energy by
looking at the projected atomic character of each eigenvalue.  The observable increase in splitting can be attributed to interplanar coupling
which is substantially suppressed when the c-axis expands to accomodate water.  Thus, the sole effect of the water on the electronic structure
is to collapse the two (nearly) concentric Fermi surfaces of the unhydrated compound until they are practically a single degenerate surface in
the hydrated compound.  While this collapse may be important, it is a purely structural effect achieved by the forced separation of Co-O planes
and is unrelated to the specific chemical composition of water.  This shows that the water itself, at least in this or similar structural
configurations, is completely irrelevant to the electronic structure of Na$_{0.33}$CoO$_2$1.33H$_2$O and that the Fermi surface is insensitive
to its presence.  This result does not depend on the specific position and orientation of the water as we obtained identical results for the
partially filled Co bands by calculating the bandstructure with water positions based on the alternate ice-like model of Ref.
\onlinecite{JWL+03}.

\begin{figure}[H] \includegraphics[width = 2.5 in]{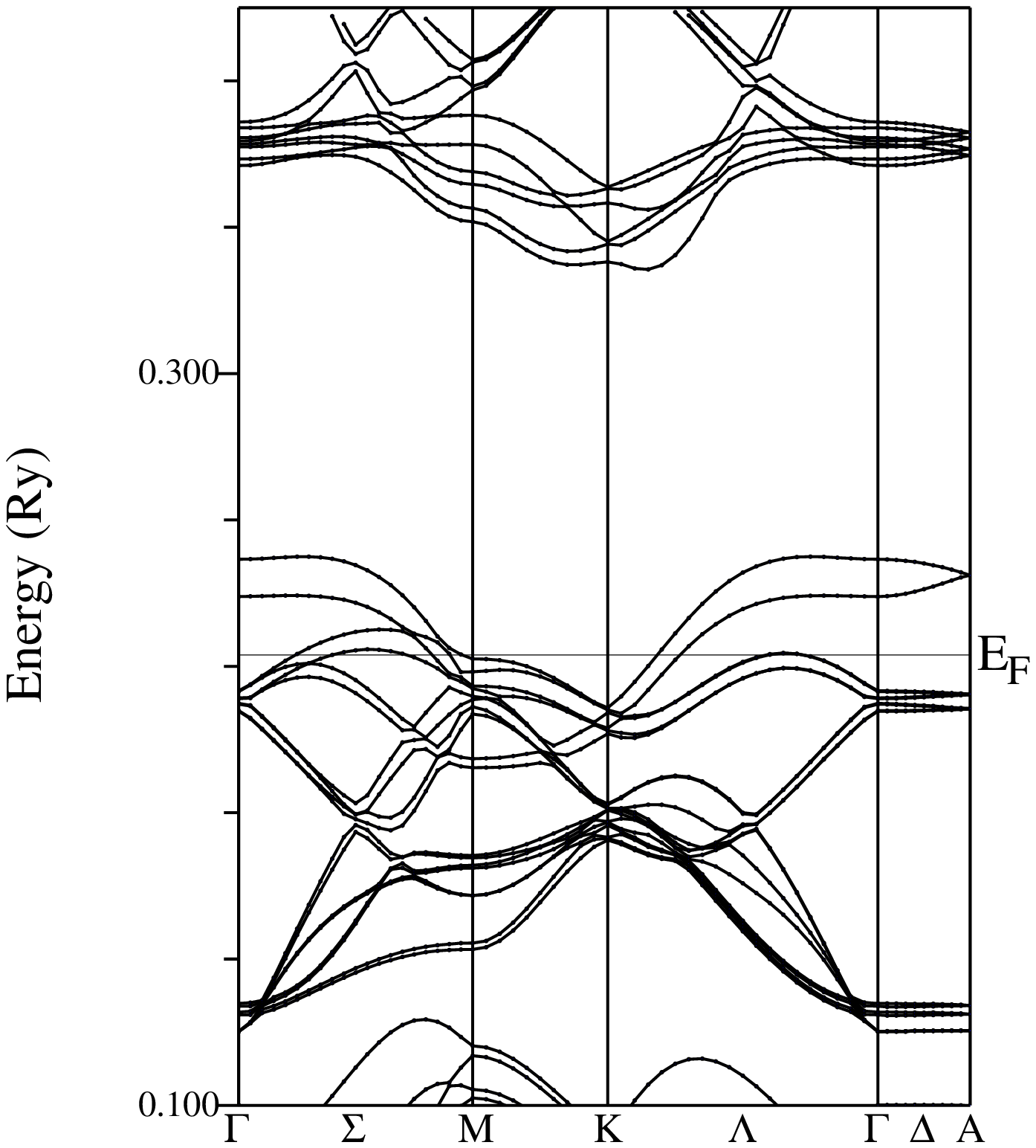} \vspace{-0.2in} \includegraphics[width = 2.5 in]{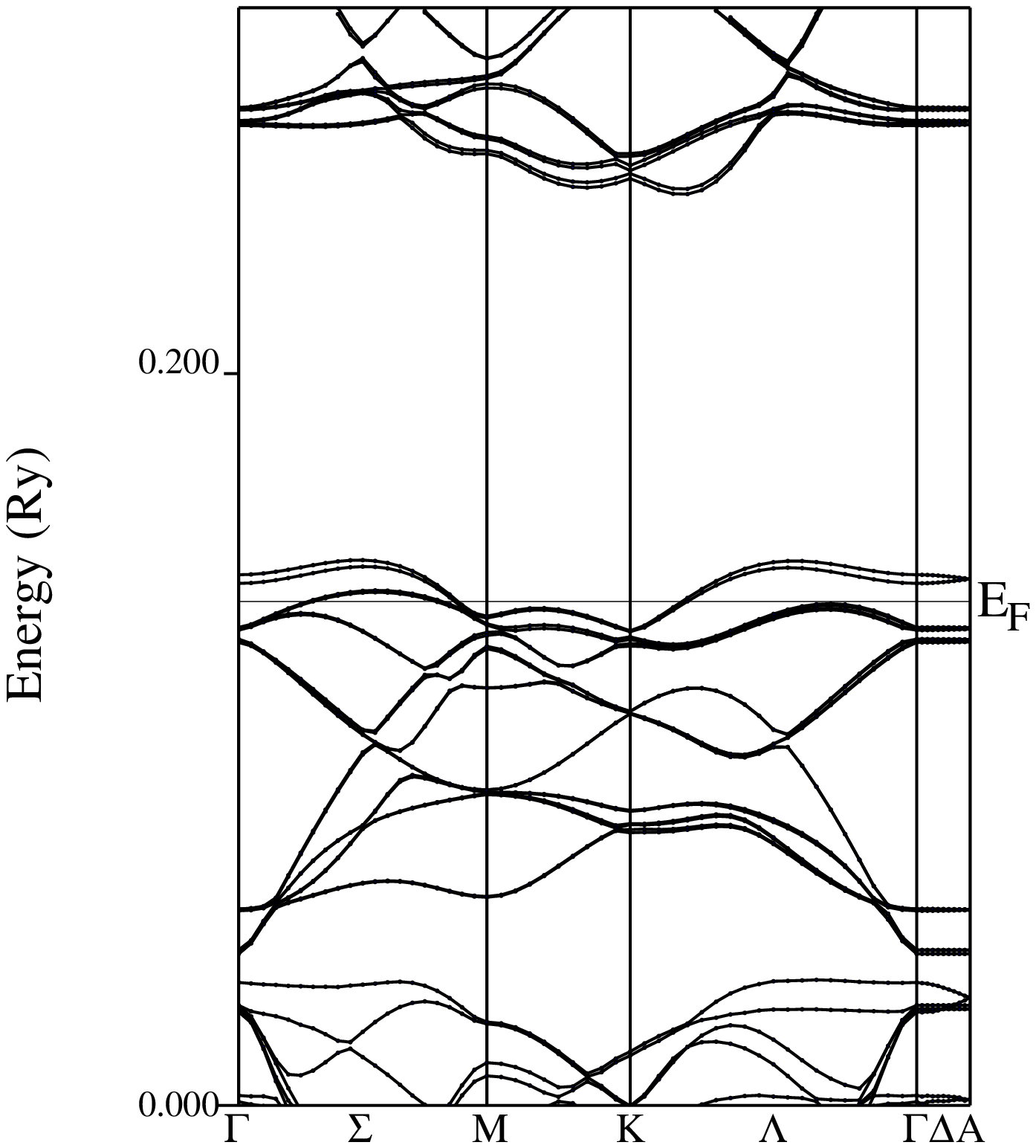}
\caption{ A comparison of the hydrated compound (right panel) with its expanded c lattice parameter and an unhydrated compound with
identical dimensions with the exception of the c-axis which remains at the unhydrated value.  The differences in dispersion between the two
structure is completely attributable to inter-planar interaction which is reduced by hydration. } \end{figure}

\section{Conclusion}
	We have shown, by explicitly including water in an LAPW calculation, that the effect of the water in Na$_{1/3}$CoO$_2$1.33H$_2$O
is overwhelmingly structural and imperceptibly electronic.  The bandstructures of the hydrated and unhydrated compounds differ only
through suppression of inter-planar coupling.  The resulting decrease in bandsplitting may have relevance to superconductivity, but the
same effect can be achieved with any spacer that sufficiently separates the Co-O planes.  The question of water's particular role in the
superconductivity is still very open, but we have shown that it has no effect on the electronic structure near the Fermi surface, other
than to make it more two dimensional.

\section{Acknowledgements}
We are grateful for helpful discussions with R. Asahi, G. Baskaran, H. Ding, T. Egami, M.Z. Hasan, W.  Koshibae, D.  Mandrus, D.A.
Papaconstantopoulos, W.E. Pickett, B.C. Sales and I. Terasaki.  M.D.J. is supported by a National Research Council Associateship.  Work at the
Naval Research Laboratory is supported by the Office of Naval Research.

\bibliography{NaCoO2.bib}

\end{document}